\documentclass[aps,preprint,amsmath,amssymb]{revtex4}
\usepackage{graphicx}
\def\be{\begin{eqnarray}}
\def\en{\end{eqnarray}}
\def\non{\nonumber}
\def\la{\langle}
\def\ra{\rangle}
\def\vp{\varepsilon}

\def\GeV{{\rm GeV} }

\begin{document}
\title{\Large \bf Productions of $K^*_0(1430)$ and $K_{1}$ in $B$ decays
}
\date{\today}
\author{\large \bf  Chuan-Hung~Chen$^{a,b}$\footnote{Email:
phychen@mail.ncku.edu.tw}, Chao-Qiang~Geng$^{c}$\footnote{Email:
geng@phys.nthu.edu.tw}, Yu-Kuo Hsiao$^{c}$\footnote{Email:
ykhsiao@phys.nthu.edu.tw } and Zheng-Tao
Wei$^{a,b,d}$\footnote{Email: weizt@phys.sinica.edu.tw }}
 \affiliation{$^{a}$Department of Physics,
National Cheng-Kung University, Tainan, 701
Taiwan\\
$^{b}$National Center for Theoretical Sciences, National Cheng-Kung
University, Tainan, 701
Taiwan\\
$^{c}$Department of Physics, National Tsing-Hua University, Hsin-Chu
, 300 Taiwan  \\
$^{d}$Institute of Physics, Academia Sinica, Taipei, 115 Taiwan}

\begin{abstract}
We study the productions of p-wave mesons $K^*_{0}(1430)$,
$K_{1}(1270)$ and $K_{1}(1400)$ in $B$ decays. By the generalized
factorization approach, we find that the branching ratios of $B\to
K^*_{0}(1430) \phi$ are similar to those of $B\to K \phi$ while
the branching ratios of $B\to K_{1}(1270) \phi$  and $B\to
K_{1}(1400) \phi$ are $O(10^{-5})$ and $O(10^{-6})$, respectively.
In terms of the observation of $B\to K_{1}(1270) \gamma$ by BELLE,
we can remove the sign ambiguity in the mixing angle for physical
states $K_1(1270)$ and $K_{1}(1400)$. In addition, we analyze
annihilation contributions in the decays $B\to K_1 \phi$ and we
conclude that they could be neglected.

\end{abstract}
\maketitle

\section{Introduction}

It is known that there have been some anomalies in penguin
dominant B decay processes, which cannot be easily explained in
the standard model (SM), especially the two puzzles: (a) the large
branching ratios (BRs) of $B\to K \eta^{\prime}$ \cite{PDG04} and
(b) the small longitudinal fractions of $B\to K^* \phi$ decays
\cite{bphik_pol}. Note that, at the quark level, both puzzles (a)
and (b) belong to the penguin dominant transitions $b\to s
q\bar{q}$. Although it is possible that some complicated hadronic
effects \cite{KS-PLB,QCD_pol} or new physics \cite{New_etaK,
New_pol}  could solve these anomalies, to find out the real causes
it is clear that we have to study more processes, in particular
those involving with similar weak interactions. Inspired by the
polarization abnormalities in $B\to K^* \phi$, we investigate the
decays of $B\to K_1 \phi$ in the SM, where $K_1$, denoting
$K_1(1270)$ and $K_1(1400)$, are axial vector bosons and the
mixtures of states $K_{^3P_{1}}$ and $K_{^1P_{1}}$. Our purpose of
this work is to see whether similar anomalies occur when these
modes are measured. Similarly, we will also study $B\to
K^*_{0}(1430) \phi$.

As usual, the challenge to study the exclusive decays is the
estimations of the transition matrix elements. By the naive
factorization (NF), the decay amplitudes can be simplified as
$c(\mu) \langle O \rangle_{\rm fact}$, in which $\langle O
\rangle_{\rm fact}$ denotes the factorizable part. Using the
approach of the NF, we immediately suffer from the problem of the
$\mu$-scale dependence on hadronic matrix elements since the
$\mu$-dependent Wilson coefficient $c(\mu)$ cannot get
compensation from $\langle O \rangle_{\rm fact}$. However, by the
QCD factorization (QCDF) \cite{QCDF} or perturbative QCD (PQCD)
\cite{PQCD} approaches, we need to know the detailed hadronic spin
structures and the associated distribution amplitudes of involving
mesons to deal with factorized and nonfactorized effects. For $B$
and $\phi$ mesons, they have been studied by the heavy quark
effective theory (HQET) \cite{KKQT} and QCD sum rules \cite{BBKT},
respectively, and at least, their asymptotic behaviors of the
leading twist and twist-3 are known clearly. Nevertheless, so far
we know nothing about the axial vector mesons of $K_{1}$. In order
to reliably estimate the relevant hadronic effects for the p-wave
modes, we employee the generalized factorization approach (GFA)
\cite{GFA1,GFA2}, in which the leading effects are factorized
parts and the nonfactorized effects
%%%%%%%%%%%%%%%%%%%%%
are lumped and characterized by the effective number of colors,
denoted by $N^{\rm eff}_c$ \cite{BSW}. Note that
 the scale and scheme
dependence on effective WCs $C^{\rm eff}_{i}$ are insensitive.

In addition, we will also analyze the annihilation contributions
which are important in $B\to PP$, $VP(PV)$ and $VV$ decays. However,
we will demonstrate that the factorized annihilation effects
in $B\to \phi K_1$
decays are smaller than those of final sates being pseudoscalars
and/or vector bosons.

The paper is organized as follows. In Sec.~\ref{sec:sec1}, we
first show the relevant effective interactions and the
parametrization of the form factors. We then give the decay
amplitudes in the framework of the generalized factorization
approach and define the polarizations for $B\to K_1 \phi$ decays.
In Sec.~\ref{sec:na}, we present our numerical analysis. We give
our conclusions in Sec.~\ref{sec:conclusion}.

\section{Form factors, decay amplitudes and polarizations}
\label{sec:sec1}

At the quark level, the effective interactions for the decays of
$B\to K^*_{0}(1430) \phi$ and $B\to K_1 \phi$ are described by
$b\to s q\bar{q}$, which are the same as $B\to K^{(*)} \phi$
decays,  and given by \cite{BBL}
\begin{equation}
H_{{\rm eff}}={\frac{G_{F}}{\sqrt{2}}}\sum_{q=u,c}V_{q}\left[
C_{1}(\mu) O_{1}^{(q)}(\mu )+C_{2}(\mu )O_{2}^{(q)}(\mu
)+\sum_{i=3}^{10}C_{i}(\mu) O_{i}(\mu )\right] \;,
\label{eq:hamiltonian}
\end{equation}
where $V_{q}=V_{qs}^{*}V_{qb}$ are the Cabibbo-Kobayashi-Maskawa
(CKM) \cite{CKM} matrix elements and the operators
$O_{1}$-$O_{10}$ are defined as
\begin{eqnarray}
&&O_{1}^{(q)}=(\bar{s}_{\alpha}q_{\beta})_{V-A}(\bar{q}_{\beta}b_{\alpha})_{V-A}\;,\;\;\;\;\;
\;\;\;O_{2}^{(q)}=(\bar{s}_{\alpha}q_{\alpha})_{V-A}(\bar{q}_{\beta}b_{\beta})_{V-A}\;,
\nonumber \\
&&O_{3}=(\bar{s}_{\alpha}b_{\alpha})_{V-A}\sum_{q}(\bar{q}_{\beta}q_{\beta})_{V-A}\;,\;\;\;
\;O_{4}=(\bar{s}_{\alpha}b_{\beta})_{V-A}\sum_{q}(\bar{q}_{\beta}q_{\alpha})_{V-A}\;,
\nonumber \\
&&O_{5}=(\bar{s}_{\alpha}b_{\alpha})_{V-A}\sum_{q}(\bar{q}_{\beta}q_{\beta})_{V+A}\;,\;\;\;
\;O_{6}=(\bar{s}_{\alpha}b_{\beta})_{V-A}\sum_{q}(\bar{q}_{\beta}q_{\alpha})_{V+A}\;,
\nonumber \\
&&O_{7}=\frac{3}{2}(\bar{s}_{\alpha}b_{\alpha})_{V-A}\sum_{q}e_{q} (\bar{q}%
_{\beta}q_{\beta})_{V+A}\;,\;\;O_{8}=\frac{3}{2}(\bar{s}_{\alpha}b_{\beta})_{V-A}
\sum_{q}e_{q}(\bar{q}_{\beta}q_{\alpha})_{V+A}\;,  \nonumber \\
&&O_{9}=\frac{3}{2}(\bar{s}_{\alpha}b_{\alpha})_{V-A}\sum_{q}e_{q} (\bar{q}%
_{\beta}q_{\beta})_{V-A}\;,\;\;O_{10}=\frac{3}{2}(\bar{s}_{\alpha}b_{\beta})_{V-A}
\sum_{q}e_{q}(\bar{q}_{\beta}q_{\alpha})_{V-A}\;,
\end{eqnarray}
with $\alpha$ and $\beta$ being the color indices.
In Eq. (\ref{eq:hamiltonian}), $O_{1}$-$O_{2}$
are from the tree level of weak interactions, $O_{3}$-$O_{6}$ are
the so-called gluon penguin operators and $O_{7}$-$O_{10}$ are the
electroweak penguin operators, while $C_{1}$-$C_{10}$ are the
corresponding Wilson coefficients (WCs). Using the unitarity
condition, the CKM matrix elements for the penguin operators
$O_{3}$-$O_{10}$ can also be expressed as $V_{u}+V_{c}=-V_{t}$.

To deal with the hadronic transition matrix elements in the
framework of the GFA, we parameterize the relevant form factors to
be
\cite{CCH} %%%%
\be \label{eq:ffp}
 \la S(p_{2})|A_\mu|\bar{B}(p_{B})\ra &=&
 -i\left[\left(P_\mu-{m^{2}_{B}-m^{2}_{S}\over q^2}\,q_
 \mu\right) F_1^{BS}(q^2)+{m^{2}_{B}-m^{2}_{S}\over q^2}
 \,q_\mu\,F_0^{BS}(q^2)\right], \non \\
 \la A(p_{2},\vp_A)|V_\mu|\bar{B}(p_{B})\ra &=&
-i\Bigg\{(m_B-m_A) \vp^*_{A\mu} V_1^{BA}(q^2)  - {\vp^*_{A} \cdot
p_{B} \over m_B-m_A} P_\mu V_2^{BA}(q^2) \non \\
&& -2m_A {\vp^*_A \cdot p_{B} \over
q^2}q_\mu\left[V_3^{BA}(q^2)-V_0^{BA}(q^2)\right]\Bigg\},
\non \\
   \la A(p_{2},\vp_A)|A_\mu|\bar{B}(p_{B})\ra &=& -{A^{BA}(q^2)\over
  m_B-m_A}\,\epsilon_{\mu\nu\rho\sigma}\vp^{*\nu}_{A} P^\rho
  q^{\sigma}, \label{saff}
 \en
with
 \be V_3^{BA}(q^2)
 &=&{m_B-m_A\over 2m_A}\,V_1^{PA}(q^2)-{m_B+m_A\over
2m_A}\,V_2^{BA}(q^2), \nonumber\\
 V_{3}^{BA}(0)&=&V_{0}^{BA}(0)\,,\nonumber
 \en
%and $V_{3}^{BA}(0)=V_{0}^{BA}(0)$,
where $S$ and $A$ denote the scalar and axial-vector mesons,
respectively, and $\vp_{A}$ is the polarization vector of the
axial vector meson. In terms of spin, orbital and total angular
momenta, they can be described by $^{2S+1}L_{J}$ so that
$S={}^{3}P_{0}$ and $A={}^{3}P_{1} ({}^{1}P_{1})$,
$P=p_{B}+p_{2}$, $q=p_{B}-p_{2}$. We note that the state $A$ is
not a physical state. Due to the decaying topology, the transition
matrix elements could be further described by
  \be
 %  X^{(BK^{*}_{0},\phi)}&=&\langle \phi|(\bar{s} s)_{V\pm A}|0\rangle \,
 %  \langle K^{*}_{0}| (\bar{s} b)_{V-A}|\bar{B}\rangle =2 \,i\, f_{\phi}m_{\phi}
 % F^{BS}_{1}(m^{2}_{\phi}) \epsilon^{*} \cdot p_{B}\nonumber, \\
  %%%%%%%%%%%%%%%%%%%%%%%%%%%%%%
   X^{(BS(A),\phi)}&=&\langle \phi|(\bar{s} s)_{V\pm A}|0\rangle \,
   \langle S(A)| (\bar{s} b)_{V-A}|\bar{B}\rangle \nonumber, \\
  %%%%%%%%%%%%%%%
   Y^{(B,\phi S(A))}_{1}&=&\langle\phi  S(A) |(\bar{q} s)_{V-A}|0\rangle \,
   \langle 0 | (\bar{q} b)_{V-A}|\bar{B}\rangle, \,\nonumber \\
   %%%%%%%%%%%%%%%%%%%%%%%
  Y^{(B,\phi S(A) )}_{2}&=&\langle \phi S(A)|(\bar{q} s)_{S+P}|0\rangle \,
   \langle 0 | (\bar{q} b)_{S-P}|\bar{B}\rangle, %\nonumber \\
   %%%%%%%%%%%%%%%%%%%%
%   X^{(B,\phi A)}_{1}&=&\langle \phi A|(\bar{q} s)_{V-A}|0\rangle \,
%   \langle 0 | (\bar{q} b)_{V-A}|\bar{B}\rangle, \,%\nonumber \\
   %%%%%%%%%%%%%%%%%%%%%%%
%   X^{(B,\phi A)}_{2}=\langle \phi A|(\bar{q} s)_{S+P}|0\rangle \,
%   \langle 0 | (\bar{q} b)_{S-P}|\bar{B}\rangle,
  \label{XY}
  \en
where $X^{(BS(A),\phi)}$ denote the factorized parts of emission
topology and $Y^{(B,\phi S(A))}_{1,2}$ stand for the factorized
parts of annihilation topology. Note that the currents  associated
with $(S+P)\otimes (S-P)$ in
 Eq. (\ref{XY})
 are from the Fierz transformations of $(V-A)\otimes (V+A)$.
From Eqs. (\ref{eq:hamiltonian})-(\ref{XY}), the decay amplitudes
for $B\to K^{*}_{0}(1430) \phi$ can be written as
 \be
  A(\bar{B}_{d}\to K^{*0}_{0}(1430) \phi)&=& \frac{G_F}{\sqrt 2}\left\{
  - V_{tb} V^{*}_{ts} \left [
  \tilde{a}^{(s)}
  X^{(BK^{*}_{0},\phi)} \right. \right. \nonumber \\
  && \left. \left.+a^{(s)}_{4} Y^{(B,\phi K^{*}_{0})}_{1} -2 a^{(s)}_{6} Y^{(B,\phi
  K^{*}_{0})}_{2}\right ]\right \}, \non
  \\
  A(B^{-}_{u}\to K^{*-}_{0}(1430) \phi)&=& \frac{G_F}{\sqrt 2}\left\{
  V_{us}V_{ub}^{*} a_{1} Y^{(B,\phi K^{*}_{0})}_{1}- V_{tb} V^{*}_{ts}
  \left [ \tilde{a}^{(s)}
  X^{(BK^{*}_{0},\phi)} \right. \right. \nonumber \\
  && \left. \left.+a^{(u)}_{4} Y^{(B,\phi K^{*}_{0})}_{1} -2 a^{(u)}_{6} Y^{(B,\phi
  K^{*}_{0})}_{2}\right ]\right \}\,, \label{eq:amp1}
 \en
 with $\tilde{a}^{(s)}=a^{(s)}_{3}+a^{(s)}_{4}+a^{(s)}_{5}$.
To be more convenient for our analysis, we can redefine the useful
WCs by combing gluon and electroweak penguin contributions to be
     \be
        a_{1}&=&C^{\rm eff}_{2}+\frac{C^{\rm eff}_1}{N^{\rm eff}_{c}},\ \ \
        a_{2}=C^{ \rm eff}_{1}+\frac{C^{\rm eff}_2}{N^{\rm eff}_{c}}, \ \
        a^{(q)}_{3}=C^{\rm eff}_{3}+\frac{C^{\rm eff}_4}{N^{\rm eff}_{c}}
        +\frac{3}{2}e_{q}\left(C^{\rm eff}_{9}+\frac{C^{\rm eff}_{10}}{N^{\rm eff}_{c}}\right),
        \nonumber \\
        a^{(q)}_{4}&=&C^{\rm eff}_{4}+\frac{C^{\rm eff}_3}{N^{\rm eff}_{c}}+\frac{3}{2}e_{q}\left(C^{\rm eff}_{10}
        +\frac{C^{\rm eff}_{9}}{N^{\rm eff}_{c}}\right),
        a^{(q)}_{5}=C^{\rm eff}_{5}+\frac{C^{\rm eff}_{6}}{N^{\rm eff}_{c}}+\frac{3}{2}e_{q}\left(C^{\rm eff}_{7}
        +\frac{C^{\rm eff}_{8}}{N^{\rm eff}_{c}}\right),\nonumber
        \\
        a^{(q)}_{6}&=&C^{\rm eff}_{6}+\frac{C^{\rm eff}_5}{N^{\rm eff}_{c}}+\frac{3}{2}e_{q}\left(C^{\rm eff}_{8}+\frac{C^{\rm
        eff}_{7}}{N^{\rm eff}_{c}}\right),
     \en
where the WCs $C^{\rm eff}_{i}$ have contained vertex corrections for
smearing the $\mu$-scale dependence in transition matrix elements
\cite{GFA2}. We note that in order to include nonfactorizable
effects, the color number $N^{\rm eff}_{c}$ is regarded as a
variable and it may not be equal to $3$. Similarly, the decay
amplitudes for $B\to A \phi$ are described by
 \be \label{eq:BphiA}
  A(\bar{B}_{d}\to A \phi)&=&\frac{G_F}{\sqrt 2}\left\{ - V_{tb} V^{*}_{ts}
  \left [ \tilde{a}^{(s)}
  X^{(BA,\phi)} +a^{(s)}_{4} Y^{(B,\phi A)}_{1} -2 a^{(s)}_{6} Y^{(B,\phi
  A)}_{2}\right ] \right \},  \non
\\
  A(B^{-}_{u}\to A \phi)&=& \frac{G_F}{\sqrt 2}\left\{ V_{us}V_{ub}^{*}
  a_{1} X^{(B,\phi A)}_{1}- V_{tb} V^{*}_{ts} \left [
  \tilde{a}^{(s)}
  X^{(BA,\phi)} \right. \right. \nonumber \\
  && \left. \left.+a^{(u)}_{4} Y^{(B,\phi A)}_{1} -2 a^{(u)}_{6} Y^{(B,\phi
  A)}_{2}\right ] \right \}. \label{eq:amp2}
 \en
As known that the physical states $K_1(1270)$ and $K_{1}(1400)$ are
the mixtures of states $^{1}P_{1}$ and $^{3}P_{1}$, their realtions
could be parametrized by \cite{Suzuki,CCH},
 \be \label{eq:mixing}
 K_1(1270)=K_{^1\!P_1}\cos\theta + K_{^3\!P_1} \sin\theta,
 \nonumber\\
 K_1(1400)=-K_{^1\!P_1}\sin\theta+K_{^3\!P_1} \cos\theta\,.
 \en
Hence, the physical decaying amplitudes are given by
 \be
  A(B\to K_{1}(1270) \phi)_{p} &=& \cos\theta \, A(B\to
  K_{^1P_1} \phi) + \sin\theta \, A(B\to
  K_{^3P_1} \phi) \nonumber \\
   %%%%%%%%%%%%%%%%%%%%%%%%%%%%%%%%
   A(B\to K_{1}(1400) \phi)_{p} &=& -\sin\theta \, A(B\to
  K_{^1P_1} \phi) + \cos\theta \, A(B\to
  K_{^3P_1} \phi). \label{eq:pamp}
  \en
%%%%%%%%%%%%%%%%%%%%%%%%%%%%%%%%%%%%%%%%%%%%%%%%%%%%%%%%%%%%%%%%%%%%%%%%%%%%%%%%%%%%%%%%

Since the final sates of $B\to AV$ carry spin degrees of freedom,  the decay amplitudes in
terms of helicities, like those in the $B\to V_{1}
V_{2}$ decays, can be generally described by
\begin{eqnarray*}
{\cal M}^{(\lambda)}
&=&\epsilon_{V\mu}^{*}(\lambda)\epsilon_{A\nu}^{*}(\lambda) \left[
a \, g^{\mu\nu} +b\;  p_B^\mu p_B^\nu + i\, c\;
\epsilon^{\mu\nu\alpha\beta} p_{1\alpha} p_{2\beta}\right]\;.
\end{eqnarray*}
Because $B$ is a pseudoscalar, the two outgoing vector mesons
$A$ and $V$ have to carry the same helicity. Consequently, the
amplitudes with different helicities can be decomposed as
\begin{eqnarray}
H_{00}&=&\frac{-1}{2m_{V}m_{A}}\left[(m^{2}_{B}-m^{2}_{V}-m^{2}_{A})a
 +2m^{2}_{B} p^{2} b\right], \nonumber\\
H_{\pm\pm}&=&a\mp m_{B} p\; c, \label{helicity}
\end{eqnarray}
where $p$ is the magnitude of vector momenta of vector mesons. In
addition, we can also write the amplitudes in terms of polarizations
as
\begin{eqnarray}
A_{L}=H_{00} \ \ \ A_{\parallel(\perp)}=\frac{1}{\sqrt{2}}(H_{--}
\pm H_{++}). \label{pol-amp}
\end{eqnarray}
Accordingly, the polarization fractions can be defined to be %
 \be
R_i=\frac{|A_i|^2}{|A_L|^2+|A_{\parallel}|^2+|A_{\perp}^2|}\,,\ \
(i=L,\parallel,\perp)\,,
\label{eq:pol}
 \en %
representing longitudinal, transverse parallel and transverse
perpendicular components, respectively. Note that $\sum_i R_i=1$.
In sum, the decay rate expressed by polarization amplitudes is
given by
 \be
  \Gamma=\frac{G^{2}_{F} p}{16\pi m^2_{B}}\left( |A_L|^2+ |A_{\parallel}|^2
  +|A_{\perp}|^2\right).
 \en

\section{Numerical Analysis} \label{sec:na}

\subsection{The analysis of annihilation contributions on $B\to AV$ decays}

It has been believed that the annihilation contributions could
significantly reduce the longitudinal polarization of $B\to K^*
\phi$ decays. It is interesting to ask whether annihilation effects
could also play an important role on the polarization fractions
of $B\to K_1 \phi$ decays.
To answer the question, we start with the analysis on the
annihilation contributions in
 $B\to PP$ and $B\to VV$ decays. For $B\to PP$
decays, the factorized amplitude associated with the
$(V-A)\otimes (V-A)$ interaction for annihilated topology can be
expressed as
\begin{eqnarray}
\langle P_{1} P_{2} | \bar{q}_{1}\gamma^{\mu}(1-\gamma_5) q_{2}\,
\bar{q}_{3} \gamma^{\mu} (1-\gamma_5) b | \bar{B} \rangle_{a}=-if_{B} (m^{2}_{1}-m^{2}_{2})F^{P_1
P_2}_{0}(m^{2}_{B})\label{eq:anni1}
\end{eqnarray}
where $m_{1(2)}$ are the masses of outgoing particles and $f_{B}$ and
$F^{P_1 P_2}_{0}(m^2_{B})$ correspond to the $B$ decay constant
 and the time-like form factor, defined
by
 \begin{eqnarray}
 \langle 0| \bar{q} \gamma^{\mu} \gamma_{5} b| \bar{B}(p_B)
\rangle &=&i f_{B} p^{\mu}_{B}\,,
\nonumber\\
 \langle P_{1}(p_1) P_{2}(p_2) | \bar{q}_{1} \gamma_{\mu} q_{2} |
 0\rangle & =& \left[q_{\mu}-\frac{m^{2}_1 -m^{2}_{2}}{Q^2}Q_{\mu}
 \right] F^{P_1 P_2}_{1}(Q^2) + \frac{m^2_1-m^2_2}{Q^2}Q_{\mu} F^{P_1
 P_2}_{0}(Q^2)\,,
\end{eqnarray}
respectively, with $q=p_1-p_2$ and $Q=p_1+p_2$. From Eq.
(\ref{eq:anni1}), it is clear that if $m_{1}= m_{2}$, the factorized
effects of annihilation topology vanish. However, if the associated
interactions are $(S+P)\otimes (S-P)$, by equation of motion, the
decay amplitude becomes
\begin{eqnarray}
\langle P_{1} P_{2} | \bar{q}_{1}(1+\gamma_5) q_{2}\, \bar{q}_{3}
(1-\gamma_5) b | \bar{B}
\rangle_{a}=if_{B}\frac{(m^{2}_{1}-m^{2}_{2}) m^{2}_{B}}{(m_{q_1} -
m_{q_2})(m_{b}+m_{q_3})} F^{P_1 P_2}_{0}(m^{2}_{B}).
\end{eqnarray}
We see that the subtracted factors appear in the numerator and
denominator simultaneously. As a result, the annihilation effects by
$(S+P)\otimes (S-P)$ interactions can be sizable due to
$(m^2_{1}-m^{2}_{2})/(m_{q_1}-m_{q_2}) \propto (m_{1}+m_{2})$. The
suppression only comes from the form factor $F^{P_1 P_2}_0(m^2_{B})
\propto 1/m^2_{B}$ which can be calculated by PQCD
\cite{LP_PLB87}.
%Following the similar discussion,
Similarly, we  expect
that the same conclusion can be given to the $VV$ modes, $i.e.$,
the longitudinal polarization should satisfy
\begin{eqnarray}
\langle V_{1L} V_{2L} | \bar{q}_{1}\gamma^{\mu}(1-\gamma_5) q_{2}\,
\bar{q}_{3} \gamma^{\mu} (1-\gamma_5) b | \bar{B} \rangle_{a}&=&
-if_{B} (m^{2}_{1}-m^{2}_{2})F^{V_1 V_2}_{0L}(m^{2}_{B}), \nonumber
\\
\langle V_{1L} V_{2L} | \bar{q}_{1}(1+\gamma_5) q_{2}\, \bar{q}_{3}
 (1-\gamma_5) b | \bar{B} \rangle_{a}&=& if_{B}
\frac{(m^{2}_{1}-m^{2}_{2}) m^{2}_{B}}{(m_{q_1} -
m_{q_2})(m_{b}+m_{q_3})} F^{V_1 V_2}_{0L}(m^{2}_{B}).
\label{eq:anni_L}
\end{eqnarray}
By the helicity analysis, we find the transverse components
to be
\begin{eqnarray} \langle V_{1T} V_{2T} |
\bar{q}_{1}\gamma^{\mu}(1-\gamma_5) q_{2}\, \bar{q}_{3} \gamma^{\mu}
(1-\gamma_5) b | \bar{B} \rangle_{a} & \propto & -i  f_{B}
(m^{2}_{1}-m^{2}_{2}) \frac{m_1 m_2}{m^{2}_{B}}
F^{V_1 V_2}_{0T}(m^{2}_{B}), \nonumber \\
\langle V_{1T} V_{2T} | \bar{q}_{1}(1+\gamma_5) q_{2}\, \bar{q}_{3}
(1-\gamma_5) b | \bar{B} \rangle_{a} & \propto & i f_{B}
\frac{(m^{2}_{1}-m^{2}_{2}) m^{2}_{B}}{(m_{q_1} -
m_{q_2})(m_{b}+m_{q_3})} \nonumber \\ && \times \frac{m_1+
m_2}{m_{B}} F^{V_1 V_2}_{0T}(m^{2}_{B}). \label{eq:anni_T}
\end{eqnarray}
Consequently, for the $VV$ modes,
 the annihilation effects of
the longitudinal polarizations by $(S+P)\otimes(S-P)$ interactions are
only suppressed by the corresponding time-like form factor $F^{V_1
V_2}_{0L}(m^2_{B})$ while those of the transverse parts are
suppressed by $(m_{1}+m_{2})/m_{B} \cdot F^{V_1 V_2}_{0T}(m^2_{B})$.
Hence, the annihilation contributions can be sizable and important on polarizations of $B\to
VV$ decays.

We now examine the decays of $B\to A V$ and check if the
suppression factor $m^{2}_{1}-m^2_{2}$ of annihilation
contributions could be smeared in the decays. Similar to the $PP$
and $VV$ cases, we start by considering the decays of $B\to SP$
with S being the p-wave scalar boson. The decay amplitude
associated with $(V-A)\otimes (V-A)$ interactions can be expressed
by
\begin{eqnarray}
\langle P_{1} S_{2} | \bar{q}_{1}\gamma^{\mu}(1-\gamma_5) q_{2}\,
\bar{q}_{3} \gamma^{\mu} (1-\gamma_5) b | \bar{B} \rangle_{a}=f_{B}
(m^{2}_{1}-m^{2}_{2})F^{P_1 S_2}_{0}(m^{2}_{B})\label{eq:anni2}.
\end{eqnarray}
By equation of motion, the decay amplitude associated
with $(S+P)\otimes (S-P)$ interactions is found to be
\begin{eqnarray}
\langle P_{1} S_{2} | \bar{q}_{1}(1+\gamma_5) q_{2}\, \bar{q}_{3}
(1-\gamma_5) b | \bar{B}
\rangle_{a}=-f_{B}\frac{(m^{2}_{1}-m^{2}_{2}) m^{2}_{B}}{(m_{q_1} +
m_{q_2})(m_{b}+m_{q_3})} F^{P_1 S_2}_{0}(m^{2}_{B}).
\end{eqnarray}
Clearly, the suppressed factor by the mass difference only appears in
the numerator, {\it i.e.} $(m^{2}_{1}-m^{2}_{2})/(m_{q_1} +
m_{q_2})\propto m_1-m_2$. As a result, we expect that the
annihilation effects in $B\to SP$ decays are much smaller than those
of $B\to PP$ decays.
%According to the results of
From Eqs. (\ref{eq:anni_L}) and (\ref{eq:anni_T}), we could
immediately see that the suppressed factor  in $B\to SP$ and $B\to
AV$ are the same. In sum, by our analysis, we conjecture that if
the final states are composed of a vector (pseudoscalar) boson and
an axial-vector (scalar) boson, the annihilation contributions
could be ignored.

Unlike $B\to SP(AV)$ decays,  there
is no extra suppressing factor for the decays of $B\to  S V$
except the $1/m^{2}_{B}$ suppression.
Nevertheless, by comparing to the dominant emission topology, due to the
$1/m^2_{B}$ suppression factor on the time-like form factor, the
annihilated effects are still small. Therefore, in our calculations
we still neglect the annihilation contributions to the BRs of $B\to
K^*_{0}(1430) \phi$.

\subsection{Branching ratios and polarization fractions}

To get the numerical estimations, we use that the decay constant
$f_{\phi}=0.233\ \rm GeV$ and the CKM matrix elements
$V_{tb}V^{*}_{ts}\approx -A\lambda^2$ with $A=0.83$ and
$\lambda=0.224$ \cite{PDG04}. Since the color number is regarded
as a variable, the effective WCs for different effective colors
are found to be $\tilde{a}^{(s)}(\mu=2.5\, \rm GeV)=(-584-97i,\,
-418-73i,\, -284-55i,\, 84-27i)\times 10^{-4}$ and
$\tilde{a}^{(s)}(\mu=4.4\, \rm GeV)=(-522-107i,\, -375-81i,\,
-257-61i,\, -80-29i)\times 10^{-4}$ for $N^{\rm eff}_{c}=(2,\,3,
5,\, \infty)$, respectively. The $\mu$-scale dependence could be
taken as  theoretical uncertainties. According to the results of
LFQCD \cite{CCH}, the values of  form factors for $B\to
K^{*}_{0}(1430)$, $B\to K_{^3P_1}$ and $B\to K_{^1P_1}$ at
$q^2=m^{2}_{\phi}$ are shown in Table~\ref{tab:vff}. %
%%%%%%%%%%%%%%%%%%%%%%%%%%%%%%%%%%%%%%%%%%%%%%%%%%%%%%%%%
\begin{table}
\caption{ \label{tab:vff} The values of form factors for $B\to
K^{*}_{0}(1430)$, $B\to K_{^3P_1}$ and $B\to K_{^1P_1}$ at
$q^2=m^{2}_{\phi}$ calculated by LFQCD \cite{CCH}. }
\begin{ruledtabular}
\begin{tabular}{cccccccc}
    & $F_1^{BK^{*}_0}$     & $V_1^{BK_{^3P_1}}$
  & $V_2^{BK_{^3P_1}}$ & $A^{BK_{^3P_1}}$
  & $V_1^{BK_{^1P_1}}$ &$V_2^{BK_{^1P_1}}$
  & $A^{BK_{^1P_1}}$ \\ \hline %
F($m^{2}_{\phi}$) &0.275&0.393&0.177&0.275&0.197& $-$0.0555&0.118%
\end{tabular}
\end{ruledtabular}
\end{table}
%%%%%%%%%%%%%%%%%%%%%%%%%%%%%%%%%%%%%%%%%%%%%%%%%%%%%%%%%%%%%%%%%%%
It is interesting to note that all values of form factors are
positive except $V^{BK_{^1P_1}}_{2}=-0.0555$. We will discuss the
implication of this negative value on the BRs and $R_{L(\perp)}$
for $B\to K_{1} \phi$. From the definition of form factors for $B$
decaying to axial-vector boson, shown in Eq. (\ref{saff}), we have
to know the masses of states $K_{^3P_1}$ and $K_{^1P_1}$. To
obtain the masses, we adopt the results of Ref. \cite{Suzuki} so
that
$m_{K_{^3P_1}}^2=m_{K_1(1270)}^2+m_{K_1(1400)}^2-m_{K_{^1P_1}}^2$
and $2m_{K_{^1P_1}}^2=m_{b_1(1232)}^2+m_{h_1(1380)}^2$. The
remaining unknown parameter is the mixing angle $\theta$. It is
known that by the decays $\tau\to \nu_{\tau} K_{1}(1270)
(K_{1}(1400))$,  $\theta$ can be determined to be around $37^0$
and $58^0$ with a twofold ambiguity \cite{Cheng-PRD67}. Recently,
$BR(B^{-}\to K^{-}_{1}(1270) \gamma)=(4.28\pm 0.94\pm 0.43) \times
10^{-5}$ has been measured by BELLE, in which the errors are
statistical and systematical, respectively. Note that there has
been no measurement on the $B^{-}\to K^{-}_{1}(1400) \gamma$ decay
yet \cite{BELLE1}. That is, the BR of $B\to K_{1}(1400)\gamma$
might be much smaller than that of $B\to K_{1}(1270)\gamma$. The
observation of the decay could remove the sign ambiguity and
conform  $\theta \approx 37^{0}$ or $58^{0}$ \cite{CC}.

\begin{figure}[htbp]
\includegraphics*[width=3.0in]{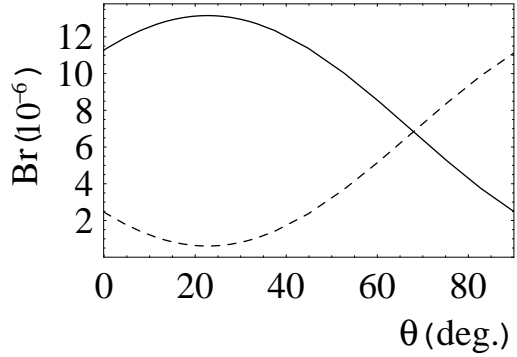}  \caption{ The braching ratios
(in units of $10^{-6})$ as a function of the mixing angle $\theta$. The
solid and dashed curves correspond to the decays of $\bar B^0\to
K_1^0(1270)\phi$ and $\bar B^0\to K_1^0(1400)\phi $, respectively. }
 \label{fig:theta}
\end{figure}

In terms of Eqs. (\ref{eq:amp1}) and (\ref{eq:amp2}), the BRs for
the different values of the mixing angle $\theta$ are displayed in
Tables~\ref{tab:br_25} and \ref{tab:br_44} with $\mu=2.5$ and
$4.4$ GeV, respectively. From the tables, we clearly see that the
BRs of $B\to (K^{*}_{0}(1430), K_{1}(1270), K_{1}(1400)) \phi$ are
increasing while $N^{\rm eff}_{c}$ is decreasing. Interestingly,
when $N^{\rm eff}_{c}=2$, $BR(B\to K^*_{0}(1430) \phi)\sim BR(B\to
K\phi)\sim 8\times 10^{-6}$
 \cite{PDG04}. It is worth
mentioning that the BRs of $B\to K\phi$ are consistent with
the data when $N^{\rm eff}_{c}=2\sim 3$ by the GFA \cite{GFA2}.
We may
conjecture that $N^{\rm eff}_{c}=2\sim 3$ is also applicable for
the decay modes with the p-wave mesons.
%According to the results of
Moreover,
from Tables~\ref{tab:br_25} and \ref{tab:br_44}, we find that if
$\theta=37^0$, $BR(B\to K_{1}(1270) \phi)$ is about one order of
magnitude larger than  $BR(B\to K_{1}(1400) \phi)$.
On the other hand,
if
$\theta=58^0$, the ratio $BR(B\to K_{1}(1270) \phi)$ to $BR(B\to
K_{1}(1400) \phi)$ is around $2$. Following the results, we
suggest that one could measure the ratio of $BR(B\to K_{1}(1270)
\phi)/BR(B\to K_{1}(1400) \phi)$ to further determine the angle $\theta$. %could be further determined.
To be more clear, we present $BR(B\to
K_1 \phi)$ with $\mu=2.5$ GeV as a function of $\theta$ in Fig.~\ref{fig:theta}. %
%%%%%%%%%%%%%%%%%%%%%%%%%%%%%%%%%%%%%%%%%%%%%%%%%%%%%%%%%%%%%%%%%%%%%
\begin{table}[htbp]
\begin{ruledtabular}
\caption{ \label{tab:br_25} The branching ratios (in units of
$10^{-6}$) of $ B\to(K_0^*(1430),\, K_1(1270),\, K_1(1400)) \phi $
decays for $\theta=37^0 (58^0)$ with $\mu=2.5\GeV$. }
\begin{tabular}{ccccc}
Mode & $N_c^{\rm eff}=2$ & $N_c^{\rm eff}=3$ &$N_c^{\rm eff}=5$ & $N_c^{\rm eff}=\infty$       \\ \hline %
$\bar B^0\to K_0^{*0}(1430) \phi$
                               & 8.06       & 4.13      &1.93      &0.18       \\ \hline %
$\bar B^0\to K_1^0(1270) \phi$ & 24.18(16.63) & 12.40(8.53) & 5.78(3.98) & 0.54(0.37) \\ \hline %
$\bar B^0\to K_1^0(1400) \phi$ &  2.66(8.70)  &  1.36(4.46) & 0.64(2.08) & 0.06(0.20) \\ \hline %
$B^-\to K_0^{*-}(1430) \phi$   & 8.77     & 4.50    & 2.10    & 0.20       \\ \hline %
$B^-\to K_1^-(1270 )\phi$      & 25.57(18.09) & 13.11(9.28) & 6.11(4.33) & 0.57(0.40) \\ \hline %
$B^-\to K_1^-(1400) \phi$      &  2.81(9.47)  &  1.44(4.85) & 0.67(2.26) & 0.06(0.21) %
\end{tabular}
\end{ruledtabular}
\end{table}
%%%%%%%%%%%%%%%%%%%%%%%%%%%%%%%%%%%%%%%%%%%%%%%%%%%%%%%%%%%%%%%%%%%%%%%%%%%%%%%%%%%
\begin{table}[htbp]
\begin{ruledtabular}
\caption{ \label{tab:br_44} The Legend is the same as
Table~\ref{tab:br_25} but  $\mu=4.4$ GeV. }
\begin{tabular}{ccccc}
Mode & $N_c^{\rm eff}=2$ & $N_c^{\rm eff}=3$ &$N_c^{\rm eff}=5$ & $N_c^{\rm eff}=\infty$       \\ \hline %
$\bar B^0\to K_0^{*0}(1430) \phi$
                               & 6.58       & 3.40      &1.61      &0.17       \\ \hline %
$\bar B^0\to K_1^0(1270) \phi$ & 19.60(13.57)& 10.14(7.01)& 4.80(3.32)& 0.50(0.35) \\ \hline %
$\bar B^0\to K_1^0(1400) \phi$ &  2.15(7.10) &  1.11(3.67)& 0.53(1.74)& 0.05(0.18) \\ \hline %
$B^-\to K_0^{*-}(1430) \phi$   & 7.16       & 3.70      & 1.75      & 0.18       \\ \hline %
$B^-\to K_1^-(1270) \phi $      & 20.73(14.76)& 10.73(7.62)& 5.08(3.61)& 0.53(0.38) \\ \hline %
$B^-\to K_1^-(1400 )\phi$      &  2.28(7.72) &  1.18(3.99)& 0.56(1.89)& 0.06(0.20) %
\end{tabular}
\end{ruledtabular}
\end{table}
%%%%%%%%%%%%%%%%%%%%%%%%%%%%%%%%%%%%%%%%%%%%%%%%%%%%%%%%%%%%%%%%%%%%%%%%%
%%%%%%%%%%%%%%%%%%%%%%%%%%%%%%%%%%%%%%%%%%%%%%%%%%%%%%%%%%%%%%%%%%%%%%%%%

\begin{figure}[htbp]
\includegraphics*[width=3.0in]{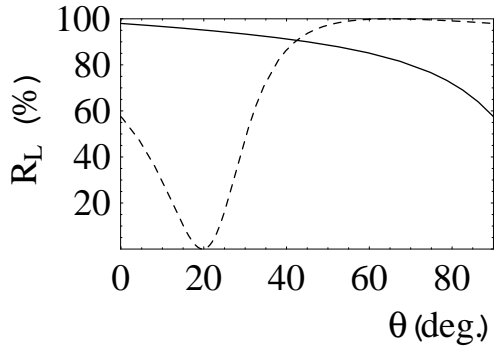}
\caption{The longitudinal
polarization fractions (in units of $\%$) as a function of the mixing
angle $\theta$. The solid and dashed curves correspond to
$\bar B^0\to K_1^0(1270)\phi$ and $\bar B^0\to K_1^0(1400)\phi $,
respectively.}
 \label{fig:pol}
\end{figure}

As discussed before, since axial-vector and vector bosons carry the
spin degrees of freedom, by the angular distribution analysis we can
study the various polarizations in $B\to A V$ decays. Hence,
from Eq. (\ref{eq:amp2}) with neglecting the annihilation
contributions, the polarization amplitudes for $B\to A$ decays are
given by
 \be
A_{L}(B\to A\phi)&=&-\frac{G_F}{\sqrt{2}}\tilde{a}^{(s)}
\frac{f_{\phi}}{2m_{A}}\left[ \left(m^{2}_{B}-m^{2}_{\phi}-m^2_{A}
\right) \left(m_{B}-m_{A}\right) V^{BA}_{1} \right. \nonumber \\
&& \left. - \frac{4m^{2}_{B} p^2}{m_{B}-m_{A}} V^{BA}_2\right]\, ,
\nonumber \\
  %%%%%%%%%%%%%%%%%%%%%%%%%%%%%%%%%%%%%%%%
A_{\parallel}(B\to A\phi)&=& G_{F} \tilde{a}^{(s)} f_{\phi} m_{\phi}
\left(
m_{B}-m_{A} \right) V^{BA}_{1}\, , \nonumber \\
  %%%%%%%%%%%%%%%%%%%%%%%%%%%%%%%%%%%%%%%%%%
A_{\perp}(B\to A\phi)&=&- G_{F} \tilde{a}^{(s)} f_{\phi} m_{\phi}
\frac{2m_{B} p}{m_{B}-m_{A}} A^{BA}\,. \label{eq:amp_pol}
 \en
The amplitudes for physical states can be obtained by following Eq.
(\ref{eq:pamp}). From the polarization amplitudes, it is clear that
by the GFA, the polarization fractions depend on the form factors
$V^{BA}_{1(2)}$, $A^{BF}$ and the mixing angle $\theta$ but they are
independent of the effective WC $\tilde{a}^{(s)}$. From Eq.
(\ref{eq:pol}) and Table~\ref{tab:vff}, our results for polarization
fractions $R_{L}$ and $R_{\perp}$ are presented in
Table~\ref{tab:pol} for $\theta=37^{0} (58^0)$. Note that
$R_{\parallel}$ can
be derived by the identity $R_{\parallel}=1-R_{L}-R_{\perp}$. %
%%%%%%%%%%%%%%%%%%%%%%%%%%%%%%%%%%%%%%%%%%%%%%%%%%%
\begin{table}
\begin{ruledtabular}
\caption{ \label{tab:pol} The polarization fractions (in unit of \%)
of $B\to ( K_1(1270),\,K_1(1400)) \phi$ with the form factors
in Table~\ref{tab:vff} and $\theta=37^{0} (58^0)$.  }
\begin{tabular}{ccc}
Mode  & $R_L$  & $R_{\perp}$  \\ \hline %
$ B\to K_1(1270) \phi$ & 91.9(85.7) &  4.2(7.8) \\ \hline %
$ B\to K_1(1400) \phi$ & 79.2(99.5) & 12.6(0.4)  %
%$B^-\to \phi K_1^-(1270)$      & 92.9(85.7) &  3.8(7.8) \\ \hline %
%$B^-\to \phi K_1^-(1400)$      & 58.3(99.5) & 24.8(0.4) %
\end{tabular}
\end{ruledtabular}
\end{table}
%%%%%%%%%%%%%%%%%%%%%%%%%%%%%%%%%%%%%%%%%%%%%%
 From Table~\ref{tab:pol}, we can see that the polarization fractions
 are somewhat insensitive to the values of  $\theta$ in $B\to
K_{1}(1270) \phi$, {\it i.e.}, $R_{L}(B\to K_{1}(1270)
\phi)=91.9 \%$ with $\theta=37^0$ while $R_{L}(B\to K_{1}(1270)
\phi)=85.7 \%$ with $\theta=58^0$. However, those for $B\to
K_{1}(1400) \phi$ are more sensitive to $\theta$, {\it i.e.}
$R_{L}(B\to K_{1}(1400) \phi)=79.2 \%$ with $\theta=37^0$ whereas
$R_{L}(B\to K_{1}(1400) \phi)=99.5 \%$ with $\theta=58^0$.
In Fig.\ref{fig:pol}, we show $R_{L}$ as a function of $\theta$.
%%%%%%%%%%%%%%%%%%%%%%%%%%%%%%%%%%%%%%%%%%%%%%%%%%%%%%%%%%%%%%%%%%%%%%%%%
%%%%%%%%%%%%%%%%%%%%%%%%%%%%%%%%%%%%%%%%%%%%%%%%%%%%%%%%%%%%%%%%%%%%%%%%%

Finally, we discuss the implication of
$V^{BK_{^1P_1}}_{2}=-0.0555$ on BRs and $R_{L(\perp)}$. In fact,
if $V^{BA}_{2}$ is positive, by comparing with $B\to K^* \phi$,
$A_{L}(B\to A\phi)$ could be smaller because the factor of
$1/(m_{B}-m_{A})$ enhances the cancellation between the two terms
in Eq. (\ref{eq:amp_pol}), whereas the corresponding factor is
$1/(m_{B}+m_{K^*})$ for $B\to K^* \phi $, which suppresses the
cancellation. However, as shown in Table~\ref{tab:pol},
$R_{L}(B\to K_1(1270) \phi)$ for $\theta=37^0$ still satisfies
$1-2m^2_{\phi}/m^2_{B}\sim O(1)$, which is the same as the
estimation for $B\to K^* \phi$ by only considering the factorized
parts. It is clear that the main reason is from the negative form
factor of $V^{BK_{^1P_1}}_{2}$. To illustrate the influence, we
tune the sign of $V^{BK_{^1P_1}}_{2}$ to be positive artificially
and we find that BRs and polarization fractions for $\theta=37^0
(58^0)$, $N^{\rm eff}_{c}=2$ and $\mu=2.5$ GeV are given as
follows:
 \be
 BR(B^0\to K^0_1(1270) \phi) &=&7.88(8.06)\times 10^{-6},
 \ \ \ BR(B^0\to K^0_1(1400)
 \phi)=0.62(0.44)\times 10^{-6} , \nonumber \\
 R_{L}(B\to K_1(1270) \phi) &=&75(69)\% , \ \ \ R_{\perp}(B\to K_1(1270)
 \phi) =13(17)\%, \nonumber \\
 R_{L}(B\to K_1(1400) \phi) &=&10(91)\% , \ \ \ R_{\perp}(B\to K_1(1400)
 \phi) =54(7)\% .
 \en
Since BRs and $R_{L}$ are reduced significantly, the measurements
on BRs and $R_{L(\perp)}$ could also test the sign of
$V^{BK_{^1P_1}}_2$.

\section{Conclusions} \label{sec:conclusion}

We have studied the productions of p-wave mesons $K^*_{0}(1430)$,
$K_{1}(1270)$ and $K_{1}(1400)$ in $B$ decays in the framework of
the GFA. In terms of form factors calculated by LFQCD, with
$N^{\rm eff}_{c}=2$ we have found that $BR(B\to K^*_{0}(1430)
\phi) \sim BR(B\to K \phi)\sim 8\times 10^{-6}$. We have also
obtained that $BR(B\to K_{1}(1270) \phi)\sim O(10^{-5})$ while
$BR(B\to K_{1}(1400) \phi)\sim O(10^{-6})$. Since the specific
values of BRs are sensitive to the mixing angle $\theta$, we can
determine the angle by the future measurements on these modes.
Moreover, we have shown that $R_{L}(B\to K_{1} \phi)\sim 80-100
\%$ and we have demonstrated that the BRs and polarization
fractions are also sensitive to the sign of the form factor
$V^{BK_{^1P_1}}_{2}$.
\\
\\
{\bf Acknowledgments}

We thank Prof. Hai-Yang Cheng and Prof. Chun-Khiang Chua for useful discussions.
This work is supported in part by the National Science Council of
R.O.C. under Grant \#s:NSC-93-2112-M-006-010 and
NSC-93-2112-M-007-014.

\end{document}